\documentclass[halfparskip,11pt]{scrartcl}

\usepackage[utf8]{inputenc}
\usepackage{amsmath}
\usepackage[standard,thmmarks]{ntheorem}
\usepackage{amsfonts}
\usepackage{mathtools}
\usepackage{tikz}
\usepackage{wrapfig}
\usepackage[numbers]{natbib}
\usepackage{mathpazo}

\usepackage{hyperref}

\usetikzlibrary{shapes.geometric}
\usetikzlibrary{calc}

\makeatletter
\let\UrlSpecialsOld\UrlSpecials
\def\UrlSpecials{\UrlSpecialsOld\do\/{\Url@slash}\do\_{\Url@underscore}}%
\def\Url@slash{\@ifnextchar/{\kern-.11em\mathchar47\kern-.2em}%
   {\kern-.0em\mathchar47\kern-.08em\penalty\UrlBigBreakPenalty}}
\def\Url@underscore{\nfss@text{\leavevmode \kern.06em\vbox{\hrule\@width.3em}}}
\makeatother

\author{Joachim Breitner\footnote{e-mail: \href{mailto:breitner@kit.edu}{breitner@kit.edu}}}
\title{Tackling the testing migration problem with SAT-Solvers}
\begin{document}
\maketitle


\begin{abstract}
We show that it is feasible to formulate the testing migration problem as a practically solvable PMAX-SAT instance, when package dependencies and conflicts are pre-processed sensibly.
\end{abstract}


\section{Introduction}

The management of software repositories such as those of Free Software distributions (Debian, Fedora,\dots) or plugin sets (Eclipse, Firefox) pose a number of interesting problems, due to dependencies and conflicts between the individual software units.

The problem discussed in this paper is that of the \emph{testing migration} that arises when preparing a release: Given a repository, containing the software that is ready to be released, and a set of newly created software, which of these may be added to the repository such that certain requirements, especially the installability of every package, are preserved. A formal definition of the problem follows in \ref{testmigdef}.

Previously, only questions related to installability have been tackled with formal methods (\cite{edos}\dots), such as which packages from a fixed repository are installable, and which packages should installed when upgrading a system. Our problem is related, but more difficult, because the installability test has to be applied to many possible choices of updated packages. This also implies that our problem at hand is $\mathcal{NP}$-hard, as testing package installability is \cite{burrows}.

The key idea in this paper is to reduce the size of the naïve but unreasonably large SAT instance by pre-processing of the interaction of dependencies and conflicts between the packages, while still leaving the hard part of the SAT solving to a general purpose SAT solver.

This approach has been implemented by the author and is deployed by the Debian project, to assist their existing, incomplement testing migration implementation. The main contributions of this paper are these:
\begin{itemize}
\item A formal description of the testing migration problem, adaptable to vaious concret applications.
\item An implementation of the testing migration problem as a PMAX-SAT instance.
\item Preprocessing steps that make this implementation practable, with correctness proofs.
\item An implementation of the solution, empirically verifying its practicability and usefulness.
\end{itemize}

\section{Background}

The setting of the testing migration problem is very similar to that of the dependency solving problem, so we extend the formalization in \cite{edos} to two repositories.

\subsection{Repositories}

The units of our problem are \emph{packages}. For these, we have an abstract set $\mathcal N$ of names and a totally ordered set $\mathcal V$ of versions. A package is a tuple of a name and a version, and $\mathcal B\subseteq \mathcal N \times \mathcal V$ the set of all packages. For a more realistical specification of the migration problem, the packages also need to carry an architecture such as i386, amd64 or arml; but these does not affect the approach described in this paper, so we ignore this aspect here.

The packages are related by a \emph{dependency function} $D \colon \mathcal B \to \mathcal P(\mathcal P(\mathcal B))$ and the \emph{conflicts relation} $C\subseteq \mathcal B \times \mathcal B$. The intended meaning of $D$ is that if $\{p_1,\ldots p_n\} \in D(p)$, then one of the $p_i$ has to be installed on a system if $p$ is to be installed. It is possible to have $\emptyset\in D(p)$; in that case the package cannot be installed. We assume here that the dependencies are already expanded: In practice, dependencies are given by a package names and version ranges. Replacing such a construct by the disjunction of all existing packages satisfying the criteria gives our dependency function, this is called \emph{dependency expansion} in \cite{edos}.

The conflicts relation is symmetric. In contrast to \cite{edos} we do not require $C$ to contain all pairs of packages with same name but different version number, but keep this relation separate:
\[
C_u \coloneqq \{(p_1,p_2) \in \mathcal B\times\mathcal B \mid \pi_1(p_1)=\pi_2(p_2) \wedge \pi_2(p_1) \ne \pi_2(p_2)\}
\]

A \emph{repository} $R\subseteq \mathcal B$ is a set of packages. An \emph{installation} $I\subseteq R$ of a repository $R$ is a selection of packages. We call the installation \emph{healthy} if all dependency and conflict relations are fulfilled: 
\[
\forall p\in I\colon  \forall d\in D(p)\colon d \cap I\ne\emptyset
\quad\text{ and }\quad
I \times I \cap C = \emptyset\,.
\]
A package $p\in R$ is \emph{installable} in $R$ if there exists a healthy installation $I\subseteq R$ with $p\in I$. A repository $R$ is called \emph{trimmed} if all its packages are installable in it.

\subsection{The testing migration problem}
\label{testmigdef}

For the formalization of the testing migration problem, we consider two repositories $T\subseteq\mathcal B$ and $U\subseteq\mathcal B$, dubbed “testing” and “unstable”. We assume that these are all packages that we need to worry about, e.g. $\mathcal B = T\cup U$. A migration is then a mo\-di\-fied testing repository $T'$ such that various requirements are fulfilled. These are in practice currently implementation-defined, the closest to a specification is in given informally in the comments in the current implementation, \texttt{britney2.py}\footnote{\url{http://anonscm.debian.org/gitweb/?p=mirror/britney2.git;a=blob;f=britney.py;hb=HEAD}}. Here, we treat these validness requirements abstractly:

\begin{enumerate}
\item (Uniqueness) A package name occurs at most once: $C_u \cap T'\times T' = \emptyset$.
\item (Trimmedness) The repository $T'$ is trimmed.
\item (Validness) Further requirements, such as that all binaries from a certain source package migrate from $T$ to $T'$ together or not at all, or that binaries where newer versions exist can only remain in $T'$ if they are from a certain section. For the purposes of this paper we just assume that $T'=T$ is always valid and that the rules can be straightforwardly formulated as a SAT instances as described in section \ref{sec:sat}.
\end{enumerate}


A choice of $T'$ is now called \emph{admissible} if all three requirements are fulfilled. We assume $T$ to be trimmed and contain every binary at most once. Therefore, $T'=T$ is always admissible, this is the \emph{trivial} migration. A migration is measured by the size of symmetric difference of $T$ and $T'$, and generally we are interested in a largest migration; but it is also of interest to find the smallest migration containing a fixed $p\in U\subseteq T$.

\subsection{SAT and PMAX-SAT}
\label{sec:sat}

Our approach to this problem is to formulate the problem as a boolean satisfiability problem (SAT), such that a solution to that problem is guaranteed to represent an admissible migration. Furthermore, if the solver allows us to mark some clauses as \emph{desired}, then we can find a largest migration; the problem then is an instance of PMAX-SAT.

Formally, given a set $V$ of atoms, an \emph{SAT instance} $P\in \mathcal P(\mathcal P(V^\pm)$ consists of sets of subsets (called clauses) of the set $V^\pm \coloneqq V \cup V^-$ of literals, where a literal is either an atom $v$ or its formal negation $v^-$. A \emph{solution} of an instance $P$ is a subset of atoms $S\subset V$, called \emph{true}, such that each clause is \emph{fulfilled}, i.e.\ has at least a true atom as a literal or a false atom as a negated literal:
\[
\forall c\in S\colon c\cap S \ne \emptyset \vee c \cap (V\setminus S)^- \ne \emptyset\,.
\]

We use $\{A \to B\}$ as an abbreviation of the clause $A^- \cup B$, where $A$ and $B$ are either set of atoms or list of atoms understood as sets, and $\{v\uparrow w\}$ as an abbreviation of $\{v^-,w^-\}$.

A \emph{PMAX-SAT instance} $P\in \mathcal P(\mathcal P(V^\pm))\times \mathcal P(\mathcal P(V^\pm))$ consits of two sets of clauses, the first set being the \emph{hard clauses} and the second set being the \emph{soft clauses}. Its solutions are those of the hard clauses, understood as a regular SAT problem, the \emph{quality} of a solution $S$ is measured by the number of fulfilled soft clauses:
\[
\#\{c\in \pi_2(P) \mid c\cap S \ne \emptyset \vee c \cap (V\setminus S)^- \ne \emptyset\}\,.
\]

For both SAT and PMAX-SAT, a variety of good general purpose solvers are available.

\section{Encoding the testing migration problem as a SAT problem}

The main ideas of this paper can be found in this section, in which we will describe an encoding of the testing migration problem as a SAT instance. We give a series of different encodings, starting with an obvious one that is incapable of handling conflicts, then a naïve, but prohibitively large encoding that handles conflicts, followed by further improvements to reduce the size of the instance.

\subsection{Encoding in absence of conflicts}

Assume first that there are no conflicts involved ($C = \emptyset$). Then the testing migration problem can be straightforwardly cast into a SAT instance. We take the set of packages as the set of atoms ($V=\mathcal B$) and define clauses that enforce the three conditions for an admissible migration:
\begin{align*}
P_u &\coloneqq \{ \{p_1 \uparrow p_2\} \mid p_1,p_2 \in C_u\} \\
P_t^1 &\coloneqq \{ \{ p \to d \} \mid p\in \mathcal B,\ d \in D(p)\}\\
P_v &\coloneqq \{ \ldots \} \\
P^1 &\coloneqq P_u \cup P^1_t \cup P_v
\end{align*}

A migration $T'\subseteq \mathcal B$ is now admissible if and only if $T'$ is a solution of $P^1$.

\begin{proof}
A solution $T'$ of $P^1$ is an admissible migration: This is obvious for the Uniqueness requirement, as it is directly expressed in the clauses in $P_u$. The (here unspecified) validness is also enforced by a straightforward representation in $P_v$.
Furthermore, $T'$ is trimmed: Every package $p\in T'$ is installable, because $I=T'$ is already a healthy installation; the dependencies are fulfilled by $P^1_t$, and there are no conflicts.

Conversely, an admissible migration $T'$ fulfills $P^1_t$: Consider a clause $c=\{p^-\}\cup d\in P^1_t$, arising from a package $p\in \mathcal B$ and a dependency disjunction $d\in D(p)$. If $p\notin T'$, then $c\cap (\mathcal B\setminus T')^- = \{p^-\} \ne \emptyset$. On the other hand, if $p\in T'$, then there exists an installation $I\subseteq T'$ with $p\in I$ and $d\cap I \ne \emptyset$, hence $d\cap T'\ne\emptyset$. So the every clause in $P^1_t$ is fulfilled.
\end{proof}

This problem encoding is sufficiently small and fast, having one atom per package under consideration. Unfortunately, it cannot be directly extended to cater for conflicts: If we take $(p_1,p_2)\in C$ to imply that $p_1 \notin T' \vee p_2 \notin T'$, then we will disallow valid migrations, as conflicts affect just installations, not repositories. So do package dependencies at a first glance, but they are positive in the sense that adding more packages to an installation does not affect the installability of existing packages negatively.

When applied to the real dataset that occurs in the migration of unstable to testing in Debian\footnote{data from 2012-03-30, 11 architectures}, this generates 263765 atoms and 1938652 clauses in $P_t$.

\subsection{Encoding with conflicts}

Now allow $C\ne \emptyset$. To represent the trimmedness of a repository directly as a SAT problem, we have to encode the search for an installation for each package. To that end, we take as atoms packags, as before, and additional atoms for each pair of packages, where we write such a pair as $p@p_i$ with the indented meaning of ``$p$ is in the installation for $p_i$'':
\[
V = \mathcal B \cup \{ p@p_i \mid p,p_i \in \mathcal B \}\,.
\]

We leave $P_u$ and $P_v$ as before and define clauses that cater for trimmedness.
\begin{align*}
P^2_e &\coloneqq \{ \{p@p_i \to p\} \mid p_i,p \in\mathcal B\} \\
P^2_i &\coloneqq \{ \{p \to p@p\} \mid p \in\mathcal B\} \\
P^2_d &\coloneqq \{ \{p@p_i \to  \{p'@p_i \mid p'\in d\}\} \mid p_i\in \mathcal B,\ p\in \mathcal B,\ d \in D(p)\}\\
P^2_c &\coloneqq \{ \{p_1@p_i \uparrow p_2@p_i\} \mid (p_1,p_2)\in C,\ p_i\in \mathcal B\}\\
P^2_t &\coloneqq P^2_e \cup P^2_i \cup P^2_d \cup P^2_c \\
P^2 &\coloneqq P_u \cup P^2_t \cup P_v
\end{align*}

A solution $S$ of this SAT instance defines a admissible repository $T' \coloneqq S\cap\mathcal B$.

\begin{proof}
For each package $p\in T'$, define an installation $I_p \coloneqq \{ p' \in \mathcal B \mid p'@p\in S\}$. This installation contains $p$ (by the clause $\{p \to p@p\}\in P^2_i$); it is a subset of $T'$ by the clauses in $P^2_e$; and it is healthy, as all dependencies of $p'\in I_p$ are fulfilled in $I_p$ by the clauses in $P^2_d$ and no conflicting package can exist in $I_p$ by the clauses $P^2_c$. So $T'$ is trimmed and, due to $P_u$ and $P_v$ as before, admissible.
\end{proof}

Conversely, for every admissible migration $T'$ there is a solution $S$ of $P^2$ such that $T' = S \cap \mathcal B$.

\begin{proof}
Since $T'$ is trimmed, we have for each package $p\in T'$ a healthy installation $I_p$; let $S \coloneqq T' \cup \{p'@p \mid p\in T',\, p'\in I_p\}$. This solution fulfills all new clauses above, as can be seen directly. The clauses in $P_u$ and $P_v$ are fulfilled as before.
\end{proof}

So we found a faithful encoding of the problem as a SAT problem instance. But it is prohibitively large if we are indeed generating variables and clauses for each pair of packages; we would be requiring 69572238990 atoms and generating 511348544780 clauses only in $P^2_d$!

\subsection{Trimming the problem}

We will reduce the size of the instance by using the first approach (encoding the dependencies directy between the variables representing the packages) when possible and fall back to the previous expensive but complete approach when required.

For that we need to introduce the “may depend” relation $\bar D$, which is an approximation of $D$:
\[
\bar D(p) = \bigcup D(p).
\]
We will most often work with its reflexive, transitive closure $\bar D^*$ and say that $\bar D^*(p)$ is the \emph{dependency closure} of $p$.

\subsubsection{Only consider possible dependencies}

It is obvious that we created way to many variables and clauses in the second attempt, as the installability of $p'$ is irrelevant when trying to find an installation for $p$ if $p'\notin \bar D^*(p)$. We phrase this as an lemma, which follows from Proposition 1 in \cite{edos}:
\begin{lemma}
If $p$ is installable in a repository $R$, then there is an installation $I$ containing $p$ such that $I\subseteq \bar D^*(p)$.
\label{instrestr}
\end{lemma}

Using this, we adjust the previous setup of the instance. The set of variables is now $\mathcal B \cup \{p'@p \mid p \in \mathcal B,\  p'\in \bar D^*(p)\}$ and the clauses are: 
\begin{align*}
P^3_e &\coloneqq \{ \{p@p_i \to p\} \mid p_i \in\mathcal B,\ p\in \bar D^*(p_i)\} \\
P^3_i &\coloneqq \{ \{p \to  p@p\} \mid p \in\mathcal B\} \\
P^3_d &\coloneqq \{ \{p@p_i \to \{p'@p_i \mid p'\in d\}\} \mid p_i\in \mathcal B,\ p\in \bar D^*(p_i),\ d \in D(p)\}\\
P^3_c &\coloneqq \{ \{p_1@p_i \uparrow p_2@p_i\} \mid p_i\in \mathcal B, (p_1,p_2)\in C|_{\bar D^*(p)}\}\\
P^3_t &\coloneqq P^3_e \cup P^3_i \cup P^3_d \cup P^3_c \\
P^3 &\coloneqq P_u \cup P^3_t \cup P_v
\end{align*}

A solution $S$ of $P^3$ is also a solution of $P^2$ and hence defines a trimmed migration.
\begin{proof}
This follows from $P^3 \subseteq P^2$.
\end{proof}

Conversely, a trimmed migration defines a solution $S$ of $P^3$ as it did for $P^2$.

\begin{proof}
By Lemma \ref{instrestr}, we can choose the installation $I_p$ of a package $p\in S\cap \mathcal B$ as a subset of $\bar D^*(p)$.
\end{proof}

This is already a considerable improvement over the naïve approach in the last section, having only 36708835 atoms and 121591516 clauses to consider.

\subsubsection{Ignore always-installable packages}

The next step is to realize that some packages $p$ have the nice property that if they are present in the repository, and installable on their own, then they can always be used to fulfill another packages dependency without worrying about $p$’s dependencies. This is trivially the case if the package has no dependencies or conflicts, but also – less trivially – if the dependency closure of $p$ does not take part in any conflicts. Let $\mathcal E \coloneqq \{ p\in \mathcal B \mid  \bar D^*(p) \cap \pi_1(C) = \emptyset \}$ be the set of these \emph{easy packages} and $\bar D_h$ the (range and domain) restriction of $\bar D$ to the hard packages $\mathcal B\setminus \mathcal E$.

This allows us to further reduce the number of atoms and clauses, by adjusting the previous setup of the instance. The set of variables is now $\mathcal B \cup \{p'@p \mid p \in \mathcal B,\  p'\in \bar D_h^*(p)\}$ and the clauses are: 
\begin{align*}
P^4_e &\coloneqq \{ \{p@p_i \to p\} \mid p_i \in\mathcal B,\ p\in \bar D_h^*(p_i)\} \\
P^4_i &\coloneqq \{ \{p \to  p@p\} \mid p \in\mathcal B\} \\
P^4_d &\coloneqq \{ \{p@p_i \to \{p'@p_i \mid p'\in d\setminus \mathcal E\} \cup \{p' \mid p'\in d\cap \mathcal E\}\} \\
&\quad\quad\quad \mid p_i\in \mathcal B,\ p \in \bar D_h^*(p_i), \ d \in D(p)\}\\
P^4_c &\coloneqq \{ \{p_1@p_i \uparrow p_2@p_i\} \mid p_i\in \mathcal B, (p_1,p_2)\in C|_{\bar D_h^*(p_i)}\}\\
P^4_t &\coloneqq P^4_e \cup P^4_i \cup P^4_d \cup P^4_c \\
P^4 &\coloneqq P_u \cup P^4_t \cup P_v
\end{align*}

Again, given a solution $S$ of $P^4$, we can construct a solution $S'$ of $P^3$.

\begin{proof}
We extend the installation of a package by all easy packages in its dependency closure:
\[
S' = S \cup \{p'@p \mid p\in S,\ p'\in \mathcal E \cap \bar D^*(p) \cap S\}.
\]
First note that an easy package $p$ with an unfulfillable dependency $\emptyset \in D(p)$ cannot be in $S$, due to $P_d^4$.
Now let $\{p@p_i \to \{p'@p_i\mid p'\in d\}\}$ be a clause in $P_d^3$ and $p@p_i\in S'$ (otherwise the clause is trivially fulfilled). If $p@p_i\notin S$, then $p\in \mathcal E \cap S$, so $d\ne \emptyset$ and there is a $p' \in d \subseteq \bar D^*(p_i)$ and $p'@p\in S$ fulfills the clause. Now assume $p@p_i \in S$. If the corresponding clause in $P^4_d$ is fullfilled by $p'@p_i\in S$ for $p' \in d\setminus\mathcal E$, this also fulfills the clause in $P_d^3$. If the clause in $P^4_d$ is fulfilled by a $p'$ with $p'\in d\cap \mathcal E$, then $p'@p_i$ is in $S'$ by definition, so the clauses in $P^3_d$ is fulfilled.

The clauses in $P_e^3$ resp. $P_c^3$ are either in $P_e^4$ resp. $P_c^4$ or have the negation of a $p@p_i$ with $p \notin \bar D_h^*(p_i)$ as a literal. As $p@p_i$ is neither in $S$ nor one of the variables added to obtain $S'$, the clauses are fulfilled.
\end{proof}

Conversely, a solution $S'$ of $P^3$ gives rise to a solution $S$ of $P^4$ by intersecting it with the variables used in $P^4$.

\begin{proof}
This holds because $P^4_e \subseteq P^3_e$, $P^4_c \subseteq P^3_c$ and clauses in $P^4_d$ correspond to clauses in $P^3_d$ with occurrences of $p'@p$ replaced by $p'$, which cannot turn from fulfilled to unfulfilled because of the clause $\{p'@p \to p'\} \in P^3_e$.
\end{proof}

This optimization reduces the size of the instance to 5235551 atoms and 31793397 clauses.

\subsubsection{Considering relevant conflicts}

The previous two refinements are subsumed by this refinement, where we identify for a package $p$ which conflicts and dependencies are actually relevant for its installability, and test the installability the verbose way only for those dependencies linking a package with its relevant conflicts.

A conflict is \emph{relevant} for $p$ if both its ends are in the dependency closure of $p$. More formally we define $C_r(p) \coloneqq C |_{\bar D^*(p)} = C\cap (\bar D^*(p) \times \bar D^*(p))$.

Now we find the connecting dependencies of a package $p$. These are those packages that are in the dependency closure of $p$ and have one end of a relevant conflict in their own dependency closure:
\[
D_r(p) \coloneqq \{p' \in \bar D^*(p) \mid \pi_1(C_r(p)) \cap \bar D^*(p') \ne \emptyset\} \cup \{p\}.
\]
We artificially add $p$ to the set to avoid special-casing this atom in $P_i$; in practice, one would omit creating an atom $p@p$ if $C_r(p)=\emptyset$.

Now we can construct the following SAT instance:
The set of variables is $\mathcal B \cup \{p'@p \mid p \in \mathcal B,\  p'\in D_r(p)\}$ and the clauses are: 
\begin{align*}
P^5_e &\coloneqq \{ \{p@p_i \to p\} \mid p_i \in\mathcal B,\ p\in D_r(p_i)\} \\
P^5_i &\coloneqq \{ \{p \to  p@p\} \mid p \in\mathcal B\} \\
P^5_d &\coloneqq \{ \{p@p_i \to \{p'@p_i \mid p'\in d\cap D_r(p_i)\} \cup \{p' \mid p'\in d\setminus D_r(p_i)\}\} \\
&\quad\quad\quad \mid p_i \in \mathcal B,\  p\in D_r(p_i),\ d \in D(p)\}\\
P^5_c &\coloneqq \{ \{p_1@p_i \uparrow p_2@p_i\} \mid p_i\in \mathcal B, (p_1,p_2)\in C|_{D_r(p_i)}\}\\
P^5_t &\coloneqq P^5_e \cup P^5_i \cup P^5_d \cup P^5_c \\
P^5 &\coloneqq P_u \cup P^5_t \cup P_v
\end{align*}

Again we transform a solution of $P^5$ into one of $P^4$ and vice-versa.

\begin{proof}
Note that easy packages are never relevant dependencies, so $D_r(p) \subseteq \bar D_h^*(p)$. By a similar argument as before, intersecting a solution $S$ of $P^4$ with the variables used in $P^5$ turns it into a solution of $P^5$.

For the other direction, let $S$ be a solution of $P^5$. Assume for this proof that the relation $\bar D$ is acyclic (otherwise, the proof is possible using fixed-point induction). For each package $p_i\in S$, define an installation recursively as
\[
I_{p_i} \coloneqq \{ p \in \mathcal B \mid p@p_i \in S \} \cup 
\bigcup \{ I_{p'} \mid p\in D_r(p_i),\ p' \in D(p),\ p'\in S,\ p'\notin D_r(p_i)\}.
\]

This installation is in the repository $S\cap \mathcal B$ prescribed by the solution: Packages from the first set are in $S$ by the corresponding clause in $P^5_e$, those from the big union because $p'\in S$ and $I_{p'}\subseteq S$ by induction. Furthermore, it contains $p_i$ because $p_i\in S$ and the corresponding clause in $P^5_i$. All packages $p\in I_{p_i}$ have their dependencies fulfilled; either because they come from an $I_{p'}$ and hence by induction, or because they come from a $p@p_i\in S$. Then $P^5_d$ ensures that each disjunction of dependencies $d\in D(p)$ contains either a $p'\in d\cap D_r(p_i)$ with $p'@p_i\in S$, hence $p'\in I_{p_i}$, or a $p' \in d\setminus D_r(p_i)$ with $p' \in S$, which would effect $I_{p'}\subseteq I_{p_i}$ and hence $p'\in I_{p_i}$

To show that $I_{p_i}$ is healthy, it remains to show that no conflict occurs in $I_{p_i}$. By the definition of $I_{p_i}$ and $D_r(p_i)$, we can see that $I_{p_i} \subseteq \bar D^*(p_i)$. Assume now that $c = (p_1,p_2) \in C\cap (I_{p_i}\times I_{p_i})$. Then $c \in C_r(p_i)$ and hence $p_1,p_2 \in D_r(p_i)$ and $p_1@p, p_2@p \in S$, which is a contradiction to the corresponding clause in $P_c^5$
\end{proof}

The numbers of this approach are 3276791 atoms and 21128454 clauses.

\section{Finding optimal solutions with PMAX-SAT}

Solving the SAT encoding described in the previous section will result in any of many possible solution, but not necessarily the best solution. Recall that migration from $T$ to $T'$ is measured by the size of the symmetric difference between $T$ and $T'$. Usually, one is interested in the largest migration. To achieve that, soft clauses are added to the problem:
\begin{align*}
P_s^{\text{max}} &\coloneqq \{ \{v\} \mid v \in U\setminus T\} \cup \{ \{v^-\} \mid v \in T\setminus U\}
\end{align*}
Feeding these together with the hard clauses $P$ from the previous section to a PMAX-SAT solver will find a solution that fulfills as many clauses from $P_s^{\text{max}}$ as possible; this number is exactly the symmetric difference between $T$ and $T'$.

Alternatively, one maybe be interested in a smallest non-trivial solution. In this case, a non-triviality clause $P_{nt}$ is added to the hard clauses, and the soft clauses are inverted:
\begin{align*}
P_{nt} &\coloneqq \{ \{ v \mid v \in U\setminus T\} \cup \{v^- \mid v \in T\setminus U\} \} \\
P_s^{\text{min}} &\coloneqq \{ \{v^-\} \mid v \in U\setminus T\} \cup \{ \{v\} \mid v \in T\setminus U\}
\end{align*}

If one is interested in a smallest migration of one particular package $p\in U\setminus T$, adding the unit clause $\{p\}$ to the hard clauses and taking $P_s^{\text{min}}$ as the soft clauses will find such a migration, if it exists. If not, then extracting the minimal unsolvable core from the SAT solver provides an explanation as to why the package does not migrate, a very helpful feature.

\section{Implementation}

Our implementation is written in Haskell and has 2300 lines of code. It can read Packages files as used by dpkg-based distributions (Debian, Ubuntu) and can generate, besides a description of final repository state, “hints” that can be fed to the currently used testing migration implementation. Therefore, it can improve the current setup without having to replace it. To detect packages that are not installable in the first place, edos-debcheck from \cite{edos} is used.

To solve SAT instances and to generate minimal unsatisfiable cores, the free SAT solver picosat \cite{picosat} is used. To solve the PMAX-SAT instances, it supports clasp \cite{clasp} and Sat4j \cite{sat4j}, the latter is used by default. We experimented with other solvers such as MiniMaxSat and MSUnCore as well, but these eliminated themselved by not being licensed under a Free Software license, a natural requirement for a project like Debian.

The code is Free Software, licensed under the GPL, and can be obtained from the code repository at \url{http://git.nomeata.de/?p=sat-britney.git}.

\section{Related work}

We build upon work done on the problem of testing the installability of packages, especially \cite{edos}. Further work in that direction investigated not only the installability of a singe package, but to termine sets of co-installable packages \cite{coinst}, and in finding good choices for upgrading an installation \cite{upgrade}, \cite{apt-pbo}.

Recent unpublished work by Jérôme Vouillon based on \cite{coinst} is also able to solve the migration problem while enforcing the stronger requirement that packages that were co-installable testing before are still co-installable afterwards. Optionally, this requirement can be relaxed, so the testing migration problem as described here can also be solved. Their implementation beforms better than ours. The main difference is that our approach finds a tractable and easily understandable encoding in SAT and uses off-the-shelve solvers, while their tool, written in OCaML, applies sophisticated transformation of the package relations, identifying equivalent packages and solving the resulting smaller problem without the help of external tools. From a users' point of view, our tool provides nothing over their tool.

%
%
%

\section{Acknowledgments}

Partially supported by the Deutsche Telekom Stiftung. I would like to thank Ralf Trainen for the invitation to Paris and also thank him, Stefano Zacchiroli, Roberto Cosmo, Mehdi Dogguy and Jérôme Vouillon for the discussion of this work.

\section{Conclusion and further work}

We have shown the feasiblity of solving the testing migration problem using off-the-shelve SAT solvers, and empirically verifyied the usefulness of the approach, applying it to the large package repositoriy created by the Debian project.

Although the current state of the program yields usable results, we expect that further reductions in the SAT problem size are possible. A considerably faster tool would allow interactive use, which can assist the distribution maintainers in finding out why a certain package does not migrate.

Currently, conflicts are considered as relevant for package which one would not expect. For example, the packages \texttt{file-rc} provides and conflicts with the common package \texttt{sysv-rc}, which appears in the transitive dependency closure of many packages (more than 8000). A sound criteria that would render such a conflict irrelevant for most packages would considerably reduce the size of the SAT instance.

Similarly, if there is a package $p'\in \bar D^*(p)$ that is \emph{independent} from $p$ in the sense that all edges leaving $\bar D^*(p')$ in the graph of dependencies and conflicts on $\bar D^*(p)$ are incident to $p'$, then any conflicts in $\bar D^*(p')$ can be removed from $C_r(p)$. It remains to be investigated if deciding this condition takes less time than is saved afterwards.

\bibliographystyle{alphadin}
\bibliography{bib}

\end{document}